\documentstyle[prl,aps,floats,twocolumn,graphics]{revtex}

\author{F. Krzakala and O.C. Martin}
\title{Spin and link overlaps in 3-dimensional spin glasses}
\address{Laboratoire de Physique Th\'eorique et Mod\`eles Statistiques,
b\^at. 100, Universit\'e Paris-Sud, F--91405 Orsay, France.}
\date{\today}
\draft

\begin{document}

\maketitle
\begin{abstract}
Excitations of three-dimensional spin glasses are computed numerically.
We find that one can flip a finite fraction of an $L~\times~L~\times~L$
lattice with an $O(1)$ energy cost, confirming the mean field picture
of a non-trivial spin overlap distribution $P(q)$. These low energy
excitations are not domain-wall-like, rather they are topologically
non-trivial and they reach out to the boundaries of the lattice.
Their surface to volume ratios decrease as $L$ increases and may
asymptotically go to zero. If so, link and window overlaps between
the ground state and these excited states become ``trivial''.
\end{abstract}


Spin glasses~\cite{Young98} are currently at the center of a hot debate.
One outstanding question is whether there exists
macroscopically different valleys whose
contributions simultaneously dominate the partition function.
At zero temperature, given the ground state configuration,
this leads one to ask whether
it is possible to flip a finite fraction of the spins and
reach a state with excess energy $O(1)$. From a mean-field
perspective~\cite{MezardParisi87b},
one expects this to be true since it happens in the
Sherrington-Kirkpatrick (SK) model. However it
is very unnatural in the context of the droplet~\cite{FisherHuse88}
or scaling~\cite{BrayMoore86} approaches where the characteristic
energy of an excitation grows with its size.
Recently it has been suggested that
the energy of an excitation may grow with its size $\ell$ as in
the droplet scaling law, $E(\ell)\approx \ell^{\theta_l}$, but only for
$\ell \ll L$, and that for $\ell=O(L)$ the energies cross
over to a different law,
$E \approx L^{\theta_g}$, where $L$ is the size of the
system~\cite{HoudayerMartin00b}.
The first exponent, $\theta_l$ ($l$ for local), may be given by
domain wall estimates, $\theta_l \approx 0.2$, while
the second exponent, $\theta_g$ ($g$ for global), could be given by the
mean field prediction, $\theta_g = 0$.
In this ``mixed'' scenario,
one has coexistence of the droplet model at finite length scales
and a mean-field behavior (if $\theta_g = 0$) for
system-size excitations ($\ell \approx L$ for which a finite fraction
of all the spins are flipped).

The purpose of this article is to provide numerical evidence that such a
mixed scenario is at work in the three-dimensional
Edwards-Anderson
spin glass. We have determined ground states
and excited states for different lattice sizes and have
analyzed their geometrical properties. The qualitative picture
we reach is that indeed $\theta_g \approx 0$.
System-size constant energy excitations are not
artefacts of trapped domain walls caused by periodic
boundary conditions, they are intrinsic to
this kind of frustrated system. The energy landscape of the
Edwards-Anderson model then consists
of many valleys, probably separated by large energy barriers.
Extrapolating to finite temperature, this picture leads to
a non-trivial equilibrium spin overlap
distribution function $P(q)$.

Given the geometric
properties of our excitations, we suggest a new scenario
for finite dimensional spin glasses: if
the surface to volume ratios of these large
scale excitations go to zero in the large $L$ limit,
then the replica symmetry
breaking will be associated with a {\it trivial} link
overlap distribution function $P(q_l)$.
We have coined this scenario TNT for
trivial link overlaps yet non-trivial spin
overlaps. Such a departure from the standard mean field picture
might hold in any dimension $d \ge 3$.

\paragraph*{The spin glass model ---}
We consider an Edwards-Anderson Hamiltonian
on a three-dimensional $L \times L \times L$ cubic lattice:
\begin{equation}
\label{eq_H_EA}
H_J(\{S_i\}) = - \sum_{<ij>} J_{ij} S_i S_j .
\end{equation}
The sum is over all nearest neighbor spins of the lattice. The
quenched couplings $J_{ij}$ are independent random variables,
taken from a Gaussian distribution of zero mean
and unit variance. For the boundaries,
we have imposed either periodic or free boundary conditions.
Although in simulations of
most systems it is best to use periodic boundary conditions so
as to minimize finite size corrections,
the interpretation of our data is simpler
for free boundary conditions. It may also be useful to note
that if boundary conditions matter in the infinite
volume limit, free boundary conditions are the experimentally
appropriate ones to use.

\paragraph*{Extracting excited states ---}
The problem of finding the ground state of a spin glass
is a difficult one. In this study we use a previously
tested~\cite{HoudayerMartin99b} algorithmic procedure
which, given enough computational ressources,
gives the ground state
with a very high probability for lattice sizes up to
$12 \times 12 \times 12$. (Since our $J_{ij}$s are continuous,
the ground state is unique up to a global spin flip.)
Our study here is limited to sizes $L \le 11$;
then the rare errors in obtaining the ground states
are far less important than our statistical errors
or than the uncertainties in extrapolating our results
to the $L \to \infty$ limit.

Our purpose is to extract low-lying excited states to see whether
there are valleys as in the mean field picture or whether the characteristic
energies of the lowest-lying large scale excitations grow with $L$ as
expected in the droplet/scaling picture. Ideally,
one would like to have a list of all the states whose excess
energy is below a given cut-off. However, because there is a
non-zero density of states associated with droplets (localized
excitations), this is an impossible task for the sizes of
interest to us. Thus instead we extract our
excitations as follows. Given the
ground state (hereafter called $C_0$), we
choose two spins $S_i$ and $S_j$ at random and force their
relative orientation to be opposite from what it
is in the ground state. This constraint can be implemented by replacing
the two spins by one new spin giving the orientation of the first spin,
the other one being its ``slave''.
We then solve for the ground state $C$ of this modified spin glass.
The new state $C$ is necessarily distinct from $C_0$ as at least
one spin ($S_i$ or $S_j$) is flipped. That flipped spin may drag along
with it some of its surrounding spins, forming a droplet of
characteristic energy $O(1)$. In the droplet picture, this
is all that happens in the infinite volume limit. However if there
exist large scale excitations with $O(1)$ energies,
then $C$ may be such
an excitation if its energy is below that of {\it all} the droplets
containing either $S_i$ or $S_j$.

\paragraph*{Statistics of cluster sizes ---}
Let $V$ be the number of sites of the cluster defining
the spins that are flipped when going from $C_0$ to $C$
(by symmetry, $V$ is taken in $\lbrack 1, L^3/2 \rbrack$).
If $P(V)$ is the probability to have an event of size $V$,
the droplet and mean field
pictures lead us to the following parametrization:
\begin{equation}
\label{eq_P_V}
P(V) = (1-\alpha) P_l(V) + \alpha P_g(V/L^3) .
\end{equation}
Here, $P_l$ and $P_g$ are
normalized probability distributions associated
with the droplet events ($V$ fixed, $L \to \infty$)
and the global events ($V = O(L^3)$).
If large scale excitations
have energies $O(L^{\theta_g})$, the ratio
$\alpha / (1 - \alpha )$ of the two contributions
should go as $L^{-\theta_g}$.
In the droplet/scaling picture, the global part
decreases as $L^{-\theta_l}$; that
is slow since $\theta_l \approx 0.2$.
In contrast, in the mean field
scenario, both the $V$ finite and the $V$ growing
linearly with $L^3$ contributions converge with non-zero weights,
$0 < \alpha < 1$, albeit with $O(L^{-\theta_l})$ finite
size corrections.

Given that the usable range in $L$ is no more than a
factor of two so that $L^{-\theta_l}$ does not vary much,
measurements of $P(V)$ on their
own are unlikely to provide stringent tests.
Nevertheless, consider the probability
$Q(v,v')$ that $V/L^3$ is in the interval $\lbrack v,v' \rbrack$.
Up to finite size corrections,
$Q(v,v') = \alpha \int_{v}^{v'} P_g(x) dx$.
In our computations, we
have used $5 \le L \le 11$, averaging for each $L$ over
2000 to 10000 randomly generated samples of the
$J_{ij}$. For each sample, we determined the ground state, and then
obtained $3$ excitations by choosing successively at random
$3$ pairs of spins ($S_i$,$S_j$).
We find that $Q(v,v')$ decreases slowly with $L$ for both
periodic and free boundary conditions, as expected in
the droplet and mean field pictures. Because
$\theta_l$ is small, when we perform fits of the form
$Q(v,v') = A + B L^{-\mu}$, we are not able
to exclude $A = 0$ nor $A \ne 0$ with any significant
confidence, so a more refined method of analysis is
necessary: we will thus consider the geometrical properties
of the events.

Before doing so, note that the statistical error on
$Q(v,v')$ depends on the number of large scale events found
in the $\lbrack v, v' \rbrack$ interval.
If the spin $S_i$ or $S_j$ has a small local field,
there is a good chance that the corresponding event will
have $V=1$, thereby reducing the statistics of the
``interesting'' events. To amplify our signal of
large $V$ events, we did not consider such spins and focused
our attention on spins in the top $25$ percentile when
ranked according to their local field. All of our data
was obtained with
that way of selecting $S_i$ and $S_j$. (Naturally, $P(V)$ and
$Q(v,v')$ depend on this choice, but the general
behavior should be the same for any choice.)

\begin{figure}
\begin{center}
\resizebox{0.7\linewidth}{!}{\includegraphics{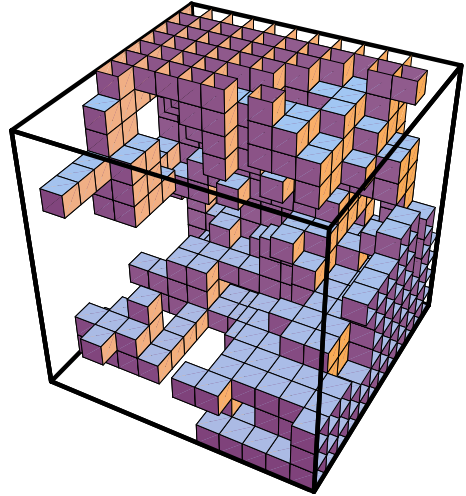}}
\end{center}
\caption{Example of excitation found for a $12^3$ lattice with
free boundary conditions.}
\label{fig_cuboid}
\end{figure}

\paragraph*{Topological features of the clusters ---}
Our claim that $\theta_g < \theta_l$ can be credible
only if our large scale excitations
are different from domain-walls
(whose energies are believed to
grow as $L^{0.2}$). It is thus useful to
consider geometrical characterizations of the excitations
generated by our procedure. Figure~\ref{fig_cuboid}
shows a typical cluster found for a $12^3$ lattice. It contains 622 spins
and its (excitation) energy is 0.98 which is $O(1)$.
The example displayed is for free boundary conditions which
permits a better visualization than periodic boundary
conditions.

The cluster shown touches
many of the 6 faces of the cube, and the same is true
for the complement of that cluster.
Such a cluster has a very non-trivial topology and is thus very far
from being domain-wall like. This motivates the following
three-fold classification of the events
we obtain when considering free boundary
conditions. In the first class, a cluster
and its complement touch all $6$ faces of the cube. In the
second class, a cluster touches at most $3$ faces
of the cube. The third class consists of all other events.
Finite size droplets should asymptotically always fall into
the second class, albeit with finite size corrections
of order $L^{-\theta_l}$.

\begin{figure}
\begin{center}
\resizebox{0.7\linewidth}{!}{\includegraphics{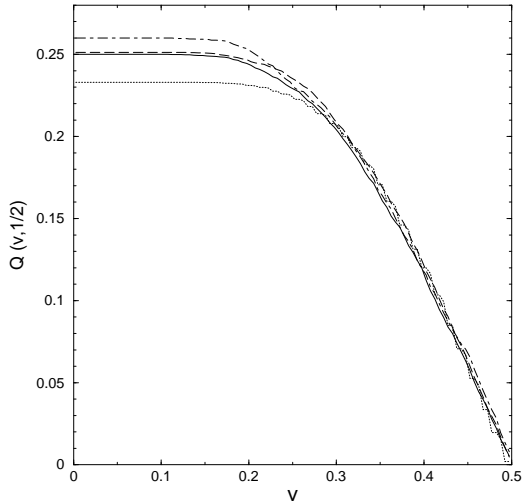}}
\end{center}
\caption{Integrated probability $Q(v,1/2)$ of events in the first
class. (From bottom to top, $L=5, 7, 9, 11$.}
\label{fig_touch}
\end{figure}

Does the first class constitute a non-zero
fraction of all events?
At finite $L$, we find the following fractions: $23.3\%$ $(L=5)$,
$23.9\%$ $(L=6)$, $25.1\%$ $(L=7)$, $24.4\%$
$(L=8)$, $25.0\%$ $(L=9)$, $25.7\%$ $(L=10)$,
and $26.0\%$ $(L=11)$.
The trend of these numbers suggests that the first class does
indeed encompass a finite fraction
of all the events when $L \to \infty$. We also considered
the scaling of cluster sizes with $L$.
Fig.~\ref{fig_touch} shows $Q(v,1/2)$
as a function of $v=V/L^3$,
restricted to events belonging to the first class.
(The $v=0$ values are the fractions
we just gave above.) The curves for different $L$ show
a small drift, $Q(v,1/2)$ {\it growing} with $L$.
We consider this drift to be a finite size effect and
that the correct interpretation of our data is
$\theta_g \approx 0$, in agreement with the mean field picture.
Our conclusion is then that
as $L \to \infty$, there is a finite probability of having an
$O(1)$ energy excitation that is non-domain-wall like,
the cluster and its complement touching {\it all} faces of
the cube.

\paragraph*{Surface to volume ratios ---}
Obviously the mean field picture obtained by extrapolating results
from the SK or Viana-Bray spin glasses
cannot teach us anything about the topology
of excitations for three-dimensional
lattices. But mean field may serve
as a guide for other properties such as the link overlap
$q_l$ between ground states and excited states. In the SK
model, the {\it spin} overlap
$q \equiv \sum S_i S_i' / N$ and the {\it link} overlap
$q_l\equiv \sum (S_i S_j)(S_i' S_j') / (N (N-1)/2)$ satisfy
$q_l = q^2$, and both $q$ and $q_l$ have non-trivial distributions.
Extrapolating this to our three dimensional system, the mean field
picture predicts that the clusters associated with large scale
excitations both span the whole system (as we saw with
free boundary conditions) and are {\it space filling}.
Quantitatively, this implies that their surface grows as the
total volume of the system, {\it i.e.}, as $L^3$.

\begin{figure}
\begin{center}
\resizebox{0.7\linewidth}{!}{\includegraphics{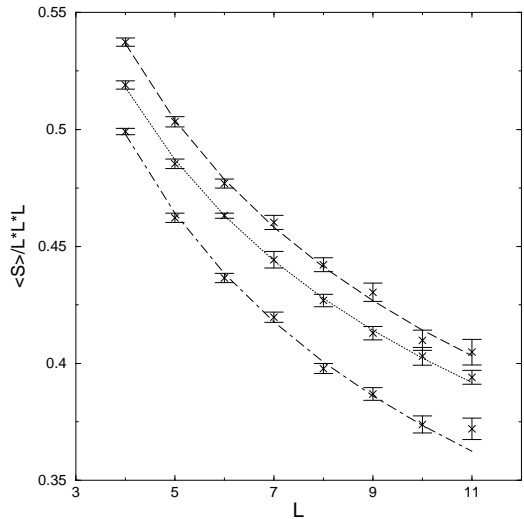}}
\end{center}
\caption{Mean value of surface to $L^3$ ratios for
$v=V/L^3$ in intervals around $0.225, 0.325, 0.425$ (bottom to top) using
free boundary conditions. Curves are pure power fits.}
\label{fig_surface}
\end{figure}

To investigate this question, we have measured the surface
of our excitations, defined as the number $S$ of links connecting
the corresponding cluster to its complement.
(Then $q_l = 1- 2 S /3 L^3$.)
In figure~\ref{fig_surface} we show the
mean value of $S / L^3$ as a function of $L$ for
$v=V/L^3$ belonging to the three intervals
$\rbrack 0.20, 0.25 \rbrack$,
$\rbrack 0.30, 0.35 \rbrack$, and
$\rbrack 0.40, 0.45 \rbrack$.
The data shown are for free boundary
conditions, but the results are very similar
for periodic boundary conditions. The most striking feature is that the
curves decrease very clearly with
$L$. For each interval, we have fitted the data to
$\langle S \rangle / L^3 = A + B/L^{\mu}$ and to a
polynomial in $1/L$. Of major interest is the value of the constant
because it gives the large $L$ limit of the curves.

Table~\ref{tab_chi2s} summarizes the quality of the
fits as given by their $\chi^2_r$ (chi squared per degree of
freedom). In all cases the fits are reasonably good; this
is not so surprizing because our range of $L$ values is small.
The most reliable fits are obtained using a quadratic polynomial in $1/L$,
this functional form leading to a smooth and
monotone behavior of the parameters
and to small uncertainties in the parameters.
For the large $L$ limits, these fits give
$A=0.22$, $A=0.27$ and $A=0.30$ for the three intervals.
(We do not give results for the
linear fits which on the contrary are very poor.) The constant
plus power fits also have good $\chi^2_r$ but the $A$s obtained were
small and {\it decreased} with $v$; also they
had large uncertainties and seemed to be compatible with $A=0$. Because
of this, we also performed fits of the form
$\langle S \rangle / L^3 = B/L^{\mu}$. These are displayed in
Fig.~\ref{fig_surface} and lead to $\mu \approx 0.30$ (the
exponent varies little from curve to curve),
again with reasonable $\chi^2_r$s. Because of this, we feel we
cannot conclude from the data that the surface to volume ratios
tend towards a non-zero asymptote. What can be said is that
this asymptote seems to be small, and that it will be difficult to
be sure that it is non-zero without going to larger
values of $L$.

\begin{table}
\begin{center}
\begin{tabular}{cccc}
Interval & $A + B/L^{\mu}$ & $A + B/L + C/L^2$ & $B/L^{\mu}$\\
\hline
$\rbrack 0.20,0.25 \rbrack$ & 0.6 & 0.6 & 2.0\\
$\rbrack 0.30,0.35 \rbrack$ & 1.1 & 1.1 & 1.5 \\
$\rbrack 0.40,0.45 \rbrack$ & 0.7 & 0.9 & 0.6 \\
\end{tabular}
\end{center}
\caption{Chi squared per degree of freedom for the fits to the data
of Fig.~\ref{fig_surface}.}
\label{tab_chi2s}
\end{table}

\paragraph*{Discussion ---}
For the three dimensional Edwards-Anderson spin glass model,
we have presented numerical evidence
that it is possible to flip a finite
fraction of the whole lattice at an energy cost of $O(1)$, corresponding
to $\theta_g \approx 0$ as predicted by mean field. This
property transpired most clearly through the use of
free boundary conditions, allowing one to conclude that
$\theta_g \approx 0$ is not an artefact of trapped domain walls caused by
periodic boundary conditions. Extrapolating to finite
temperature, we expect the equilibrium $P(q)$ to be non trivial as in the
mean field picture.

The other messages of our work concern the nature of
these large scale excitations whose energies are $O(1)$. First,
using free boundary conditions, we found them to be
topologically highly non-trivial:
with a finite probability they reach the boundaries
on all 6 faces of the cube. Thus they are not domain-wall-like,
rather they are sponge-like. Second, our data (both for
periodic and free boundary conditions) indicate very clearly that
their surface to volume ratios decrease as $L$ increases. The
most important issue here is whether or not these ratios decrease to zero
in the large $L$ limit. Although our data are
compatible with a non-zero limiting value as predicted
by mean field, the fits were not conclusive
so further work is necessary.

If the surface to volume ratios turned out to go to zero, we would
be lead to a new scenario that we have coined ``TNT''. In the standard
mean field picture, the surface to volume ratios cannot go
to zero; indeed
in the SK and Viana-Bray spin glass models
there are {\it no} spin clusters with
surface to volume ratios going to zero.
However, in finite dimensions, one can have
surface to volume ratios going to zero, in which case
$q_l \to 1$. This property would then lead to a non-trivial
$P(q)$ but to a trivial $P(q_l)$.
This trivial-non-trivial (TNT)
scenario does not seem to have been proposed previously.

Perhaps the most dramatic consequence of this new scenario is for window
overlaps in spin glasses: because in TNT one is asymptotically always
in the bulk of an excitation,
correlation functions at any finite distance will show
no effects of replica symmetry breaking. That this may arise in
fact is supported by work by Palassini and
Young\cite{PalassiniYoung99a}
who showed that certain window overlaps seemed to become
trivial as $L \to \infty$.
(See also~\cite{Middleton99} for a similar discussion in
two-dimensions.) These authors
referred to this property as evidence for
a ``trivial ground state structure''.
But in our picture the {\it global} (infinite distance)
structure is non-trivial, as indicated by
$\theta_g = 0$, in sharp contrast to the droplet/scaling
picture. Also, in very recent work~\cite{YoungPrivate},
Palassini and Young have extended their previous investigations
and have extracted excited states by a quite different
method from ours, and
they find that their data is compatible with
the TNT scenario.
Naturally, there is also evidence in {\it favor}
of the non-triviality
of window overlaps~\cite{MarinariParisi99b}.
Nevertheless, we believe that our mixed scenario is
a worthy candidate to describe the physics of short range
spin glasses. Furthermore, its plausibility should not
restricted to $3$ dimensions, it could hold in
all dimensions greater than $2$. (Note that in $d=2$, excitations
are necessarily topologically trivial.) An important indication
of this was obtained by Palassini and Young whose
computations~\cite{YoungPrivate} favor the TNT
scenario over the droplet picture
in the 4-dimensional Edwards-Anderson model.


\paragraph*{Acknowledgements ---}
We thank J.-P. Bouchaud and M. M\'ezard for very stimulating
discussions and for their continuous encouragement, and
M. Palassini and A.P. Young for letting us know about their
work before publication. Finally, we thank
J\'er\^ome Houdayer; without his superb work on the genetic
renormalization approach~\cite{HoudayerMartin99b}, this numerical
study would not have been possible.
F.K. acknowledges support from the
MENRT, and O.C.M. acknowledges support from the Institut Universitaire de
France. The LPTMS is an Unit\'e de Recherche de
l'Universit\'e Paris~XI associ\'ee au CNRS.

\bibliographystyle{prsty}
\bibliography{../../../Bib/references}

\addcontentsline{toc}{chapter}{\protect\bibname}
\begin{thebibliography}{10}

\bibitem{Young98}
{\em Spin Glasses and Random Fields}, edited by A.~P. Young (World Scientific,
  Singapore, 1998).

\bibitem{MezardParisi87b}
M. M{\'e}zard, G. Parisi, and M.~A. Virasoro, {\em Spin-Glass Theory and
  Beyond}, Vol.~9 of {\em Lecture Notes in Physics} (World Scientific,
  Singapore, 1987).

\bibitem{FisherHuse88}
D.~S. Fisher and D.~A. Huse, Phys. Rev. B {\bf 38},  386  (1988).

\bibitem{BrayMoore86}
A.~J. Bray and M.~A. Moore,  in {\em Heidelberg Colloquium on Glassy Dynamics},
  Vol.~275 of {\em Lecture Notes in Physics}, edited by J.~L. van Hemmen and I.
  Morgenstern ({S}pringer, {B}erlin, 1986), pp.\ 121--153.

\bibitem{HoudayerMartin00b}
J. Houdayer and O.~C. Martin, Euro. Phys. Lett. {\bf 49},  794  (2000).

\bibitem{HoudayerMartin99b}
J. Houdayer and O.~C. Martin, Phys. Rev. Lett. {\bf 83},  1030  (1999).

\bibitem{PalassiniYoung99a}
M. Palassini and A.~P. Young, Phys. Rev. Lett. {\bf 83},  5126  (1999).

\bibitem{Middleton99}
A. Middleton, Phys. Rev. Lett. {\bf 83},  1672  (1999).

\bibitem{YoungPrivate}
M. Palassini and A.~P. Young, private communication.

\bibitem{MarinariParisi99b}
E. Marinari {\it et~al.}, J. Stat. Phys. {\bf 98},  973  (2000).

\end{thebibliography}

\end{document}